\documentclass[twocolumn,showpacs,preprintnumbers,amsmath,amssymb]{revtex4}

\usepackage{graphicx}
\usepackage{dcolumn}
\usepackage{bm}

\newcommand{\bea}{\begin{eqnarray}} 
\newcommand{\beq}{\begin{equation}} 
\newcommand{\ear}{\end{array}} 
\newcommand{\eea}{\end{eqnarray}} 
\newcommand{\eeq}{\end{equation}}

\begin{document}

\title{\Large Band terminations in density functional theory.}

\author{A.\ V.\ Afanasjev} 

\address{Department of Physics and Astronomy, Mississippi State
University, MS 39762, USA}

\date{\today}

\begin{abstract}

   The analysis of the terminating bands has been performed in the relativistic
mean field framework. It was shown that nuclear magnetism provides an additional 
binding to the energies of the specific configuration and this additional binding 
increases with spin and has its {\it maximum} exactly at the terminating state. 
This suggests that the terminating states can be an interesting probe of the 
time-odd mean fields {\it provided that other effects can be reliably isolated.}
Unfortunately, a reliable isolation of these effects is not that simple: many
terms of the density functional theories contribute into the energies of the terminating 
states and the deficiencies in the description of those terms affect the result. 
The recent suggestion \cite{ZSW.05} that the relative energies of the terminating 
states in the $N \neq Z,\ A\sim 44$  mass region given by $\Delta E$ {\it provide 
unique and reliable constraints on time-odd mean fields and the strength of 
spin-orbit interaction} in density functional theories has been reanalyzed. The 
current investigation shows that the $\Delta E$ value is affected also by the 
relative placement of the states with different orbital angular momentum
${\it l}$, namely, the placement  of the $d$ (${\it l}=2$) and $f$ (${\it l}=3$) 
states. This indicates the dependence of the $\Delta E$ value on the properties
of the central potential.
\end{abstract}

\pacs{PACS: }

\maketitle

\section{Introduction}

 The density functional theory (DFT) in its non-relativistic 
\cite{BHR.03} and relativistic \cite{VRAL} realizations is a standard 
tool of modern nuclear structure studies. However, providing global 
description of atomic nuclei, it  still suffers from the fact that 
many channels of effective interaction are not uniquely defined: 
this is a reason for a large variety of the DFT parametrizations, 
the quality of many of which is poorly known. The spin-orbit interaction  
and the time-odd mean fields are of particular interest in 
this context, since there are considerable variations for these 
quantities (see, for example, Refs.\ \cite{DD.95,BRRMG.99,AR.00}).  
The spin-orbit interaction plays a crucial role in the definition of 
the shell structure of nuclei, and, thus its accurate description
is required so that theoretical tools have predictive power for nuclei 
beyond known regions. The time-odd mean fields  (or nuclear magnetism 
(NM) in the language of the relativistic mean field (RMF) theory \cite{KR.90}) 
contribute to the single-particle Hamiltonian only in situations where the 
intrinsic time-reversal symmetry is broken and Kramers degeneracy of 
time-reversal counterparts of the single-particle levels is removed. 
The rotating nuclei and odd and odd-odd mass nuclei are typical 
examples of such situations, see Refs.\ \cite{BHR.03,VRAL} and 
references quoted therein.

  It was recently suggested in Ref.\ \cite{ZSW.05} that the set of 
terminating states in the $N \neq Z,\ A\sim 44$  mass region {\it provides 
unique and reliable constraints on time-odd mean fields and the strength 
of spin-orbit interaction} in Skyrme density functional theory (SDFT), 
see also Refs.\ \cite{S.07,ZS.07}. Later this procedure (called as 
'TS-method' in this manuscript, where 'TS' refers to 'terminating 
states') has been used in the analysis of terminating states in this mass 
region within the RMF theory \cite{BWSMG.06} 
which is one of the versions of covariant density functional (CDFT) 
theory  \cite{VRAL}. 

  The authors of Refs.\ \cite{ZSW.05,S.07,ZS.07,BWSMG.06} 
claim that the TS-method is free from the drawbacks of standard methods 
of defining spin-orbit interaction based on measuring the single-particle 
energies of the spin-orbit partner orbitals in spherical nuclei. As a 
consequence, it is stated that it allows to define very accurately 
both isoscalar and isovector channels of spin-orbit interaction 
\cite{ZSW.05,BWSMG.06}; the feature which was impossible in the 
previously existing methods.

   The conclusions obtained within the TS-method are drastically different 
from the ones previously obtained in the SDFT and RMF frameworks. For example, 
based on the comparison of the calculated and experimental energies of 
spin-orbit partner orbitals, it was shown in Ref.\ \cite{BRRMG.99} that the 
experimental spin-orbit splittings are better reproduced in the RMF approach 
than in the SDFT (see Fig.\ 2 in Ref.\ \cite{BRRMG.99}). On the contrary, the 
results obtained in Refs.\ \cite{ZSW.05,BWSMG.06} within the TS-method show 
that the SDFT provides better description of spin-orbit splittings than the 
RMF: it was suggested in Ref.\ \cite{ZSW.05} that only 5\% reduction in 
isoscalar {\it ls} strength is needed in the SDFT approach in order to 
reproduce experimental data. Considering the conflict of these results 
and the importance of the spin-orbit interaction in nuclei it is necessary 
to understand to which extent the basis of the suggested TS-method is sound 
and justified.  The goal of the present manuscript is the study of the 
properties of terminating bands and their terminating states in the RMF 
framework.  In particular,  the question of whether all DFT contributions
have been correctly accounted in the realization of the TS-method in Refs.\ 
\cite{ZSW.05,BWSMG.06} is addressed is the current manuscript.

The manuscript is organized as follows. Time-odd mean fields in terminating
bands are studied in Sect.\ \ref{Ne20-sect}. The basis of the TS-method, 
its realization in self-consistent DFT and in the Nilsson potential are 
discussed in Sect.\ \ref{TS-method}. Sect.\ \ref{Th-vs-exp} analyses the 
contributions of different DFT terms into the relative energies of 
terminating states in the $A\sim 44$ mass region.  The discussion of the 
energy scale, its connection to the effective mass of the nucleon 
and their impact on the relative energies of terminating states is 
presented in Sect.\ \ref{energy-scale}. Finally, Sect.\ \ref{concl} 
summarizes main conclusions.

\section{Time-odd mean fields in terminating bands: test case of
$^{20}$Ne}
\label{Ne20-sect}

 Previous DFT investigations of the modifications of the moments of inertia
\cite{KR.93,CRMF,DD.95,YM.00} and single-particle properties \cite{AR.00} in 
rotating nuclei caused by the time-odd mean fields (nuclear magnetism) were 
restricted to the superdeformed (SD) bands. However, these bands are far away from 
the termination and are characterized by a relatively stable deformation. In 
order to understand how NM affects the properties of the terminating bands, the 
ground state configuration in $^{20}$Ne has been studied. 
This band is a classical example of band termination \cite{CNS}. It has the
$\pi (d_{5/2})^2_4 \nu (d_{5/2})^2_4$ configuration relative to the $^{16}$O core
with maximum spin $I_{max}=8^+$. The selection of this configuration has been 
guided by its simplicity,  which allows us to understand the role of time-odd 
mean fields in terminating bands in greater details. Although the terminating 
bands were observed also in heavier nuclei, it is difficult to trace them from 
low spin up to band termination in the self-consistent approaches \cite{VRAL} 
without special techniques such as used in the cranked Nilsson-Strutinsky (CNS)
approach \cite{CNS}.

\begin{figure}
\centering
\includegraphics[width=8.0cm]{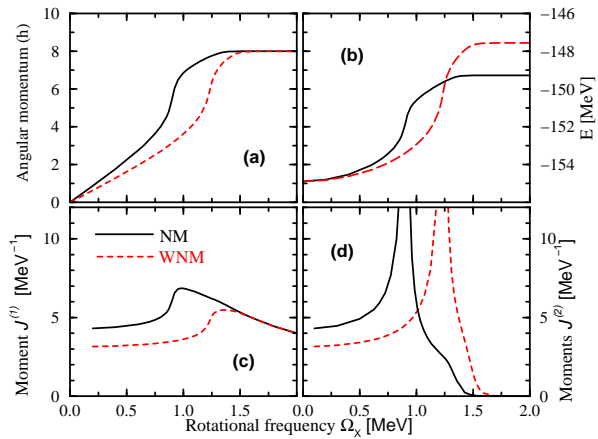}
\caption{(Color online)  Calculated total angular momentum (panel (a)), total binding 
energy (panel (b)), kinematic and dynamic moments of inertia (panels (c) 
and (d)) as a function of rotational frequency $\Omega_x$ in the ground 
state band of $^{20}$Ne. The results obtained with and without nuclear 
magnetism  are presented. Note that the band termination takes place at 
rotational frequency above which the subsequent increase of rotational
frequency does not modify neither total angular momentum nor total 
binding energy. This point also corresponds to $\gamma=60^{\circ}$ 
(Fig.\ \ref{ne20-spin}c).
\label{ne20-freq}}
\end{figure}

    The investigation of NM in $^{20}$Ne has been carried out within the 
framework of the cranked relativistic mean field (CRMF) theory \cite{KR.89,CRMF} 
following the formalism of  Ref.\ \cite{AR.00}, where similar study has been
performed for the yrast SD band in $^{152}$Dy. In the CRMF calculations of 
this manuscript, all 
fermionic and bosonic states belonging to the shells up to $N_F=12$ and 
$N_B=16$ are taken into account in the diagonalization of the Dirac equation 
and the matrix inversion of the Klein-Gordon equation, respectively. The detailed 
investigation indicates that this truncation scheme provides very good numerical 
accuracy. The  NL1 \cite{NL1} parametrization of the RMF Lagrangian is employed 
in the study of $^{20}$Ne, while the studies of terminating states in the $A\sim 44$ 
mass region (Sects.\ \ref{TS-method} and \ref{Th-vs-exp}) are performed mostly 
with the NL1 and NL3 \cite{NL3} parametrizations. The pairing is neglected in 
calculations.

  Fig.\ \ref{ne20-freq} shows the total 
angular momentum, total binding energy $E$ and kinematic (J$^{(1)}$) and 
dynamic (J$^{(2)}$) moments of inertia of the ground state configuration 
in $^{20}$Ne as a function of rotational frequency obtained in the 
calculations with and without NM (the later will be further denoted as 
WNM). The band crossing caused by the interaction of the $r=+i$ 
signatures of the [220]1/2 and [211]3/2 orbitals both in the proton and 
neutron subsystems takes place at lower frequency in the calculations 
with NM; this is in line with previous finding that the NM shifts the 
band crossing frequencies \cite{AFR.02}. The NM also increases the 
moments of inertia before band crossing (Fig.\ \ref{ne20-freq}c,d); 
similar effect has been seen before in the SD bands (see Ref.\ 
\cite{AR.00} and references therein). It also leads to a faster 
alignment of angular momentum with rotational frequency (Fig.\ 
\ref{ne20-freq}a); full alignment at $I_{max}=8^+$ corresponding to 
band termination  takes place at lower frequency in the calculations with NM.

\begin{figure}
\centering
\includegraphics[width=8.0cm]{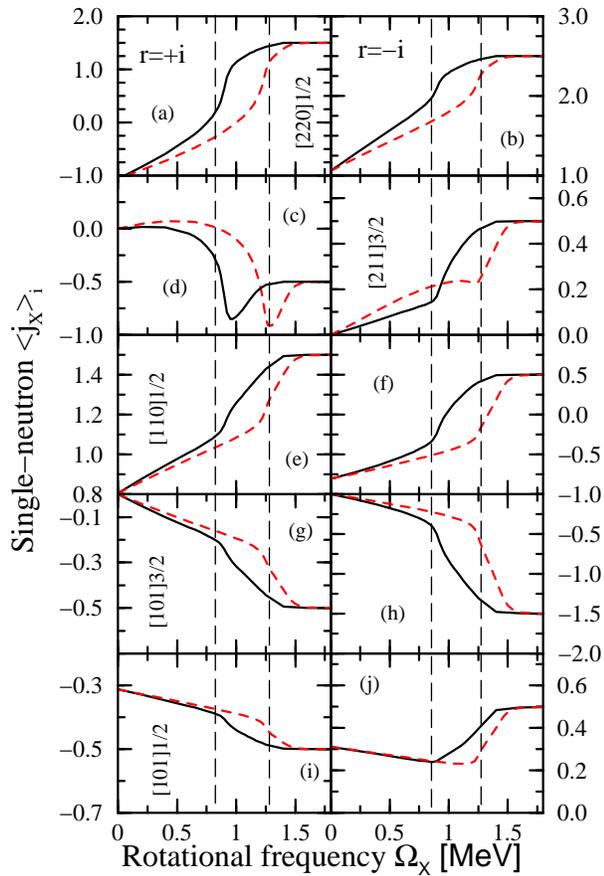}
\caption{ (Color online) The expectation values of single-particle angular 
momenta $<j_x>_i$ of the neutron orbitals occupied at low frequency in the 
ground state configuration of $^{20}$Ne given along the deformation path of this 
configuration. The single-particle orbitals are labeled by means of 
the asymptotic quantum numbers
$[Nn_z\Lambda]\Omega$ (Nilsson quantum numbers) of their dominant
component of the wave function at $\Omega_x=0.0$ MeV. The orbitals
with signature $r=+i$ and $r=-i$ are shown in right and left panels,
respectively. The results of the calculations with and without NM
are shown by solid and dashed lines, respectively. The region of
the band crossing is located between the dashed lines.
\label{fig-ne20-jx}}
\end{figure}

\begin{figure}
\centering
\includegraphics[width=8.0cm]{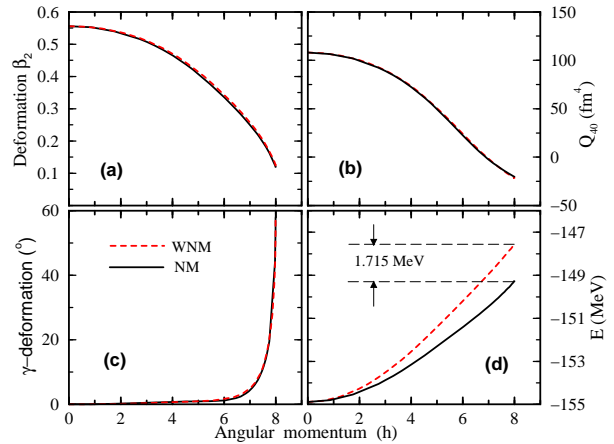}
\caption{Calculated quadrupole deformation $\beta_2$ (panel (a)), mass 
hexadecapole moment $Q_{40}$ (panel (b)), $\gamma$-deformation (panel 
(c)), and total binding energy $E$ (panel (d)) as a function of
angular momentum. The results obtained with (NM) and without (WNM)
nuclear magnetism are presented.
\label{ne20-spin}}
\end{figure}

 In the context of study of terminating states two results are 
important. First, at the band termination the NM does not modify neither 
total angular momentum (Fig.\ \ref{ne20-freq}a) nor the expectation values 
of the single-particle angular momenta $<j_x>_i$ (Fig.\ \ref{fig-ne20-jx}). 
At lower frequency, the impact of 
NM on $<j_x>_i$ is similar to the one previously studied in the SD band 
of $^{152}$Dy \cite{AR.00}, and, thus, it will not be discussed in detail. 
However, one should mention that when analyzing the impact of NM on $<j_x>_i$, 
the region of band crossing and the region close to the band termination have 
to be excluded from consideration because considerable differences in the 
deformations of the NM and WNM solutions at given frequency distort their 
comparison.
Second, the NM provides an additional binding to the energies
of the specific configuration and this additional binding increases with
spin and  has its maximum 
exactly at the terminating state (Fig.\ \ref{ne20-freq}b and \ref{ne20-spin}d)). 
This suggests that the terminating states can be an interesting probe of 
the time-odd mean fields {\it provided that other effects can be reliably 
isolated.}

   When the results of the NM and WNM calculations are compared as a function 
of total angular momentum, one can see that the quadrupole deformation 
$\beta_2$ (Fig.\ \ref{ne20-spin}a), mass hexadecapole moment $Q_{40}$ (Fig.\ 
\ref{ne20-spin}b), and $\gamma$-deformation (Fig.\ \ref{ne20-spin}c) are 
almost the same in both calculations. The only difference is seen in the total 
binding energies (Fig.\ \ref{ne20-spin}d), where the NM solution is more bound 
than the WNM solution. These results give a hint why  the 
cranked models based on the phenomenological potentials like Woods-Saxon or 
Nilsson, which do not include time-odd mean fields \cite{DD.95}, are so successful in the 
description of experimental data. When considered as a function of spin  
the deformation properties of the rotating  system are only weakly affected 
by the time-odd mean fields, and the proper renormalization of the moments of 
inertia \cite{CNS} takes care of the $E$ versus angular momentum curve.

\section{The TS-method}
\label{TS-method}

  The TS-method suggested in Ref.\ \cite{ZSW.05} employs the terminating 
states in the $A\sim 44$ mass region with the proton $d_{3/2}^{-1}\,f_{7/2}^{n+1}$ 
and $f_{7/2}^{n}$ structure, but, in general, according to Ref.\ \cite{ZS.07} 
can be employed to terminating states in any mass region. The difference 
$\Delta E^{exp}= E(d_{3/2}^{-1}\,f_{7/2}^{n+1})-E(f_{7/2}^{n})$ between the excitation 
energies $E(d_{3/2}^{-1}\,f_{7/2}^{n+1})$ and $E(f_{7/2}^{n})$ of the terminating 
states with the structure $d_{3/2}^{-1}\,f_{7/2}^{n+1}$ and $f_{7/2}^{n}$ is dominated 
by the size of the magic gap 20 which is surrounded by the $d_{3/2}$ and $f_{7/2}$ 
spherical subshells (see Fig.\ \ref{method}).

\subsection{The spin-orbit splittings in the TS-method}

\begin{figure}[h]
\includegraphics[angle=0,width=8cm]{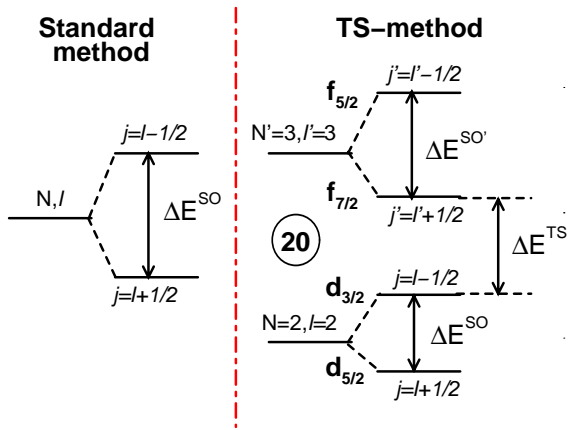}
\caption{\label{method} Schematic comparison of the standard and 
TS-methods of defining the strength of spin-orbit interaction. In the
left and right panels, the single-particle spectra without (on left) 
and with (on right) spin-orbit interaction are compared. }
\end{figure}

 The principal difference between the standard and the TS-methods of 
defining the strength of spin-orbit interaction is schematically shown 
in Fig.\ \ref{method}. The standard method requires that both partners 
of spin-orbit ${\it ls}$ doublet with $\it j=l\pm \frac{1}{2}$ are observed 
in experiment since the spin-orbit splitting $\Delta E^{SO}$ is related 
to the strength of spin-orbit interaction. This severely restricts the 
possibilities to study spin-orbit interaction since both partners should 
be located in the vicinity of the Fermi level to be observed: this 
condition is very difficult to satisfy for high-$j$ orbitals since they 
are characterized by large spin-orbit splittings, see, for example, Refs.\ 
\cite{BRRMG.99,IEMSF.02}. On the contrary, the TS-method employs the 
terminating states based on the particle-hole excitations involving the
single-particle states with $\it j=l-\frac{1}{2}$ and $\it j'=l'+\frac{1}{2}$
which emerge from different $N$-shells and are characterized by the energy 
splitting $\Delta E^{TS}$ (Fig.\ \ref{method}). 
Since these states are located in the vicinity of the Fermi level, the TS-method 
provides also information on the  spin-orbit interaction of high-$j$ orbitals 
according to Ref.\ \cite{ZSW.05}.

  It is necessary to recognize that both methods of defining the spin-orbit interaction
are not free from important drawbacks. The experimental single-particle states in 
spherical nuclei used in the standard method  are strongly affected  
by the couplings with vibrations in many cases \cite{MBBD.85}. On the other hand, the 
$\Delta E^{TS}$ value used in the TS-method depends not only on the spin-orbit 
splitting but also on how well the positions of the single-particle states with 
different orbital momenta $l$ and $l'$ (Fig.\ \ref{method}) are described in the DFT 
calculations.  The later fact has been neglected in Ref.\ \cite{ZSW.05} using 
an analogy with the Nilsson potential, the validity of which is questioned below.

\subsection{The TS-method in self-consistent approaches}
\label{TS-SC}
  
   In the self-consistent calculations, the 
$\Delta E^{SC}= E(d_{3/2}^{-1}\,f_{7/2}^{n+1})-E(f_{7/2}^{n})$ quantity
is defined as the difference of the binding energies of the corresponding 
terminating states. Without going into the details of specific DFT 
(nonrelativistic or relativistic), one can conclude that 
$\Delta E^{SC}$ depends on
\begin{itemize}
\item 
the energy scale of the single-particle spectra which is related to the 
effective mass $m^*(k_F)/m$ of the nucleon  at the Fermi surface,

\item 
the spin-orbit interaction,

\item 
the relative placement of states with different angular momentum $l$,

\item
the time-odd mean fields (see Sects.\ \ref{Ne20-sect} and \ref{NM-A44}),

\item
the polarization effects (both in time-even and time-odd channels) on 
going from the $f_{7/2}^{n}$ to the $d_{3/2}^{-1}\,f_{7/2}^{n+1}$ terminating 
state.
\end{itemize}
For simplicity of discussion, they will occasionally be called as 
'ingredients of $\Delta E^{SC}$'. 

  $\Delta E^{SC}$ can also be split into the terms which depend 
on time-even (TE) and time-odd 
(TO) mean fields
\begin{eqnarray}
\Delta E^{SC}=\Delta E^{SC}_{TE}-\Delta E^{SC}_{TO}
\end{eqnarray}
The minus sign in front of $\Delta E^{SC}_{TO}$ reflects the fact 
that NM always decreases the size of $\Delta E$.

\subsubsection{Coupling constant dependence of $\Delta E^{SC}$}
\label{coupl-const}

  The ingredients of $\Delta E^{SC}$ depend in a complicated 
way on different terms of the DFT with at least one term 
contributing into each of four first ingredients of $\Delta E^{SC}$ 
within the nonrelativistic SDFT. In the RMF theory, the spin-orbit 
interaction is defined in a natural way without additional coupling 
constant \cite{VRAL}. The time-odd mean fields related to NM are 
defined through the Lorentz covariance \cite{VRAL} and also do not 
require an additional coupling constant. However, both these terms 
depend in an indirect way on the coupling constants of other terms 
of the RMF Lagrangian. 

  Considering the uncertainties of the description of the 
'ingredients of $\Delta E^{SC}$' in the DFT, it is not obvious that 
simple fit (to experimental $\Delta E^{exp}$ values) of the coupling 
constants of the DFT terms related to a pair of ingredients of 
$\Delta E^{SC}$ (such as time-odd mean fields and spin-orbit 
interaction as in Ref.\ \cite{ZSW.05}; the later treated perturbatively) 
will allow to define these constants in a unique way. This is especially 
true considering that physical observables depend on many (or sometimes all) 
coupling constants simultaneously within the DFT, and the effect of varying 
one or two coupling constants may be either enhanced or cancelled by a 
variation of others. Strictly speaking, the quality of such perturbative 
fits involving only one or two terms of the DFT is not known until the results 
of global fit including all the DFT terms are available. For example, it was 
shown in Ref.\ \cite{LBBDM.07} that perturbative studies of tensor terms 
allow only very limited conclusions.

\subsubsection{Polarization effects}
\label{pol-eff}

\begin{figure}[h]
\includegraphics[angle=0,width=8cm]{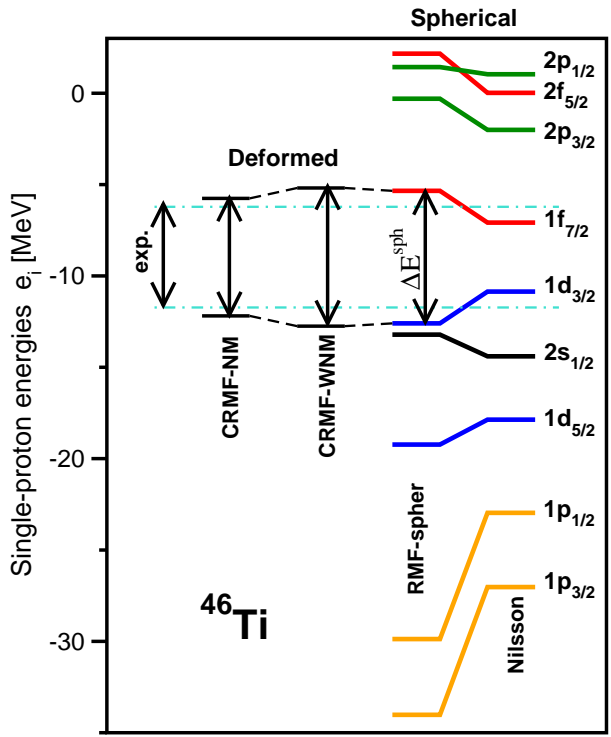}
\caption{\label{single} (Color online) (right part) Single-particle energies  
at spherical shape in $^{46}$Ti obtained in the RMF calculations 
with the NL1 parametrization and the Nilsson potential with standard set 
of parameters \cite{BR.85} (columns indicated as ``RMF-spher'' and 
``Nilsson''). In order to facilitate the comparison, the RMF states are 
given at the calculated energies, while the energies of the states 
obtained in the Nilsson potential are shifted in such a way that the 
average energy  of the $1f_{7/2}$ and $1d_{3/2}$ states is the same in both 
calculations. (left part) The magnitudes $\Delta E$ of the $f_{7/2}-d_{3/2}$ 
splittings  as extracted from the energies of terminating states 
($\Delta E = E(d_{3/2}^{-1}\,f_{7/2}^{n+1})-E(f_{7/2}^{n})$) are shown by 
arrows in columns ``exp'' (experiment), ``CRMF-NM'' and ``CRMF-WNM'' 
(CRMF calculations with and without NM). The middle points of arrows are 
located at the average energy  of the $1f_{7/2}$ and $1d_{3/2}$ RMF states. 
The terminating states are deformed in the CRMF calculations.
Note that the value of $\Delta E= 5.1$ MeV (not shown in figure) obtained 
in the cranked Nilsson-Strutinsky calculations compares favorably 
with experiment ($\Delta E^{exp}= 5.51$ MeV).}
\end{figure}

\begin{figure}[h]
\includegraphics[angle=-90,width=8cm]{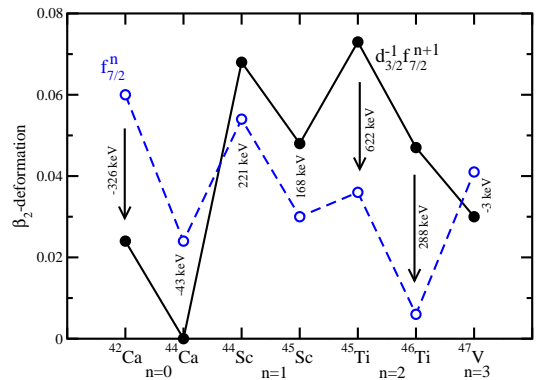}
\caption{\label{def} (Color online) Quadrupole deformations of 
the terminating  states obtained in the CRMF calculations with the 
NL1 parametrization. For each nucleus, the deformation polarization 
energies $\Delta E^{def-pol}$ are shown between the calculated 
deformation points. Three nuclei with the largest $\Delta E^{def-pol}$ 
values are indicated by arrows.}
\end{figure}

  Fig.\ \ref{single} illustrates the polarization effects present in the CRMF 
calculations of terminating states. Since fully stretched states with spin 
$I_{max}$ are reasonably well described by a single Slater determinant \cite{ZSNSZ.07}, 
the comparison with experiment is performed without angular momentum restoration 
as in all DFT studies of the terminating $I_{max}$ states in this mass region,
see, for example, Refs.\ \cite{ZSW.05,BWSMG.06}. At spherical shape, the 
$f_{7/2}-d_{3/2}$ energy gap ($\Delta E^{sph}$) 
in the single-particle 
spectra considerably exceeds $\Delta E^{exp}$. The $\Delta E^{sph}$ value is a 
very good approximation to the results of the spherical RMF calculations without 
NM in which the gap between these states is defined as the difference of the 
binding energies of the $d_{3/2}^{-1}\,f_{7/2}^{n+1}$ and $f_{7/2}^{n}$ states: the 
difference between two results for all nuclei under study does not exceed 40 keV. When 
deformation polarization effects denoted as $\Delta E^{def-pol}$ are taken into account 
(the ``CRMF-WNM'' column in Fig.\ \ref{single}), this gap becomes even larger and 
reaches $\Delta E= 7.57$ MeV exceeding by 37\% the experimental 
value. The inclusion of NM decreases the difference between experiment and 
calculations considerably (by 1.14 MeV) (column ``CRMF-NM'' in Fig.\ \ref{single}).
Note that mass and charge quadrupole and hexadecapole moments change only by 
$\sim 10^{-4}\%$ on going from the CRMF-WNM to CRMF-NM solutions. Thus, the deformation 
differences between these two solutions are almost non-existant and can be neglected. 
Based on the consideration of polarization effects, the $\Delta E^{SC}$ can be 
approximated as
\begin{eqnarray}
\Delta E^{SC} \approx \Delta E^{sph} + \Delta E^{def-pol}-\Delta E^{SC}_{TO}
\label{eq-poleff}
\end{eqnarray}
Considering that the terminating states of interest are close to spherical
(Fig.\ \ref{def}), this approximation which corresponds to a perturbative 
treatment of the deformation polarization effects should be quite reasonable. 
This approximation also allows to use the results of spherical RMF calculations 
in subsequent analysis of $\Delta E$ (Sect.\ \ref{Th-vs-exp}). 

\subsection{The Nilsson potential analogy of Ref.\ \cite{ZSW.05}}
 
  In order to overcome the problems discussed in Sect.\ \ref{coupl-const},
the authors of Ref.\ \cite{ZSW.05} use the analogy with the simple form 
of the Nilsson potential \cite{N.55} for the $\Delta E^{SC}_{TE}$ part.
This form of the Nilsson potential is similar to the one given below
(Eq.\ (\ref{Nil-compl})), but with the parameters $\kappa$ and $\mu$ independent on 
principal quantum number $N$. However, no proof (analogy is not a proof) is 
provided whether such a transition from self-consistent DFT to the Nilsson 
potential is valid and whether the dependence of the energies of the 
single-particle states on quantum numbers $N, j, l,$ and $s$ is the same 
in the self-consistent DFT and in the Nilsson  potential. 
Fig.\ \ref{single} clearly shows that the later is not a case.

 In the simple form of the Nilsson potential, the magnitude of the $f_{7/2}-d_{3/2}$ 
splitting related to the magic gap 20 is given by
\begin{eqnarray}
\Delta E^{Nil}= \hbar \omega_0 (1-6\kappa - 2 \kappa \mu). 
\label{Delta-Nil-simple}
\end{eqnarray}
Thus, it depends on three major factors: (i) the energy scale of 
the single-particle potential characterized by $\hbar \omega_0$, 
(ii) the flat-bottom and surface properties entering through the 
orbit-orbit term $\sim \mu$, 
and (iii) the strength of the spin-orbit term $\kappa$. Then, the authors 
of Ref.\ \cite{ZSW.05} using the fact that in light nuclei the Nilsson 
potential resembles the pure harmonic oscillator potential, which 
leads to $\mu \sim 0$, conclude that the magnitude of the 
$f_{7/2}-d_{3/2}$ splitting is given by
\begin{eqnarray}
\Delta E^{Nil} = \hbar \omega_0 (1-6\kappa). 
\label{Delta-Nil-0}
\end{eqnarray}
Thus, in this approximation, the $\Delta E^{Nil}$ splitting is defined only by 
the energy scale $\hbar \omega_0$ and the strength of the $\it{ls}$-potential. 
Arguing that the energy scale is rather well constrained by the data not only 
for the Nilsson but also for the self-consistent approaches, the authors of 
Ref.\ \cite{ZSW.05} conclude that the $f_{7/2}-d_{3/2}$ splitting is directly 
related to the strength of the $\it{ls}$-potential. Or alternatively,
this approximation corresponds to the situation when the $\Delta E^{Nils}$
does not depend on orbital motion of the nucleons.

\subsection{An alternative form of the Nilsson potential}
\label{alternative}

 One question of paramount importance we have to ask is whether the 
simple form of the Nilsson potential given in Ref.\ \cite{N.55} and used 
in Ref.\ \cite{ZSW.05} is unique and how well it describes the 
experimental data. It turns out that the modern versions of the Nilsson 
potential employ the parameters $\kappa$ and $\mu$ which are dependent 
on the main oscillator quantum number $N$ and on nucleon type 
(proton or neutron), thus facilitating the study of wide range of nuclei 
with the same set of single-particle parameters and with comparable accuracy 
\cite{BR.85,Ragbook,Rag-priv}. Since the Nilsson potential is phenomenological
in nature, this procedure is well justified. The accuracy of the description of 
different physical quantities such as, for example, rotational properties and 
relative energies of different single-particle configurations 
\cite{BR.85,Ragbook,CNS,GBI.86} is  considerably improved when different 
values of $\kappa$ and $\mu$ are used for different $N$-shells. Although some 
variations between the parametrizations exist, this approach is used in almost 
all parametrizations of the Nilsson potential developed from the middle of 
the 80ties of the last century \cite{GBI.86,BR.85,A150}.  The studies employing this 
description of the Nilsson potential are abundant and provide systematic 
information on the accuracy of the description of physical observables in 
different mass regions. Even more sophisticated dependence of the $\kappa$ 
and $\mu$ parameters on the principal quantum number $N$ and orbital angular 
momentum $l$ is introduced in Ref.\ \cite{S.86} and employed in a number of 
studies, see, for example, Refs.\ \cite{PRet.97,BZ.95}.

\subsubsection{Terminating states in the $A\sim 44$ mass region}

 In the Nilsson potential with $\kappa_N$ and $\mu_N$ parameters dependent on 
the principal oscillator quantum number $N$ \cite{BR.85,Ragbook} 
\begin{eqnarray}
&&\hat{H}^{Nil-N-dep}  -  \frac{3}{2}\hbar \omega_0  =  \nonumber \\
 & = & \hbar \omega_0 \{N-\kappa_N [2 {\bm l}{\bm s}
+\mu_N (\bm{l}^2 - <\bm l^2 >_N)]\},
\label{Nil-compl}
\end{eqnarray}
the magnitude of the $f_{7/2}-d_{3/2}$ splitting related to the energy 
difference of the $d_{3/2}^{-1}\,f_{7/2}^{n+1}$ and $f_{7/2}^{n}$
terminating states is given by
\begin{eqnarray}
&&\Delta E^{Nil-N-dep}= \\ \nonumber
&=& \hbar \omega_0 (1- 3[\kappa_2+\kappa_3] - 3 \kappa_3 \mu_3 
- \kappa_2 \mu_2). 
\end{eqnarray}
The superscript $'Nil-N-dep'$ is used to indicate the dependence of 
these expressions on the main oscillator quantum number $N$. 

\begin{table}
\caption{ The standard parametrization of the Nilsson potential from Ref.\ 
\protect\cite{BR.85,Ragbook}. Only the parameters for the $N=2$ and 3 
 shells are shown.}
\begin{center}
\begin{tabular}{|c|c|c|c|c|} \hline
      &  \multicolumn{2}{|c|}{Protons} &  \multicolumn{2}{|c|}{Neutrons} \\
 N    &  $\kappa$ & $\mu$    &  $\kappa$  &  $\mu$       \\ \hline
 2    &  0.105    &  0.00    &  0.105     &  0.00        \\ 
 3    &  0.090    &  0.30    &  0.090     &  0.25        \\ \hline
 \end{tabular}
\end{center}
\label{Table-par}
\end{table}

  The  $\kappa_N$ and $\mu_N$ parameters of the so-called standard parametrization 
of the Nilsson potential are shown for the shells of interest in Table \ref{Table-par}. 
For the protons, the value $\Delta E^{Nil-N-dep}=0.415 \hbar \omega_0$ is obtained in 
the calculations employing the $\kappa_N$ parameters from the Table \ref{Table-par} 
but assuming the $\mu_N=0$  as it was done in the derivation of Eq.\ (\ref{Delta-Nil-0}). 
This value can be compared with the $\Delta E^{Nil-N-dep}=0.334 \hbar \omega_0$ value 
obtained with the $\kappa_N, \mu_N$ parameters from the Table \ref{Table-par}. One can 
see that these two values differ by approximately 25\% and this difference is solely 
attributed to the non-zero value of the $\mu_3$ parameter. Considering that 
$\Delta E^{exp}\sim 5.5$ MeV (see Fig.\ \ref{excit}),  25\% difference correspond 
to 1.4 MeV; this difference definetely cannot be ignored when experimental data is 
compared with experiment.

\subsubsection{Concluding remarks}

 Even for terminating states in the $A\sim 44$ mass region one 
cannot ignore the dependence of the energies of the {\it (N,l)} and  
{\it (N',l')} states, from which the ${\it j=l-\frac{1}{2}}$ and 
${\it j'=l'+\frac{1}{2}}$ states used in the TS-method emerge (see Fig.\ 
\ref{method}), on the orbital angular momentum. This dependence enters 
through the $\mu_N (\bm{l}^2 - <\bm l^2 >_N)]$ term of the Nilsson 
potential (Eq. (\ref{Nil-compl})). This is contrary to the approximation made 
in the derivation of Eq.\ (\ref{Delta-Nil-0}) which has a consequence 
that the energy difference $\Delta E^{Nils}$ depends only on the energy 
scale $\hbar \omega_0$ and the strength of the spin-orbit term $\kappa$. 
The dependence of $\Delta E^{Nils}$ on the orbital angular momentum in 
the case of terminating states involving single-particle states from 
higher $N$-shells has been recognized in Ref.\ \cite{ZS.07}.

\section{Terminating states in the $A\sim 44$ mass region: what we can learn 
from the comparison with experiment}
\label{Th-vs-exp}

\begin{figure}[h]
\vspace{0.5cm}
\includegraphics[angle=0,width=8cm]{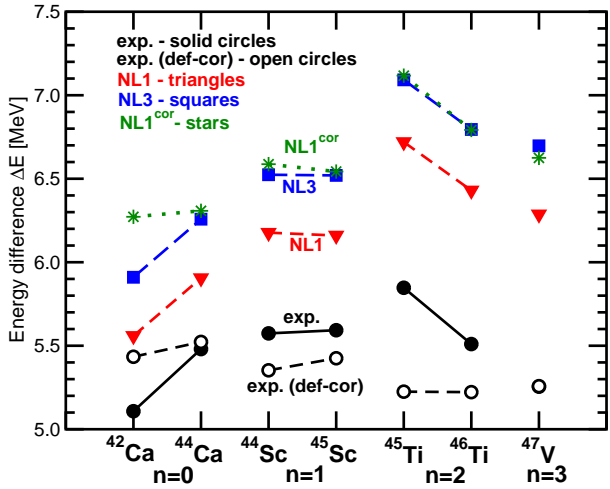}
\caption{\label{excit} (Color online) The experimental and calculated 
magnitudes $\Delta E$ of the $f_{7/2}-d_{3/2}$ splittings  as extracted 
from the energies of terminating states 
($\Delta E = E(d_{3/2}^{-1}\,f_{7/2}^{n+1})-E(f_{7/2}^{n})$). The results of 
the CRMF calculations are shown for the NL1 and NL3 parametrizations of 
the RMF Lagrangian. The experimental data corrected for the deformation 
polarization effects (as obtained in the NL1 parametrization) is shown
by open circles. Note that the experimental and deformation polarization
corrected values of $\Delta E$ coincide in the case of $^{47}$V.}
\end{figure}

\begin{figure}[h]
\includegraphics[angle=0,width=8cm]{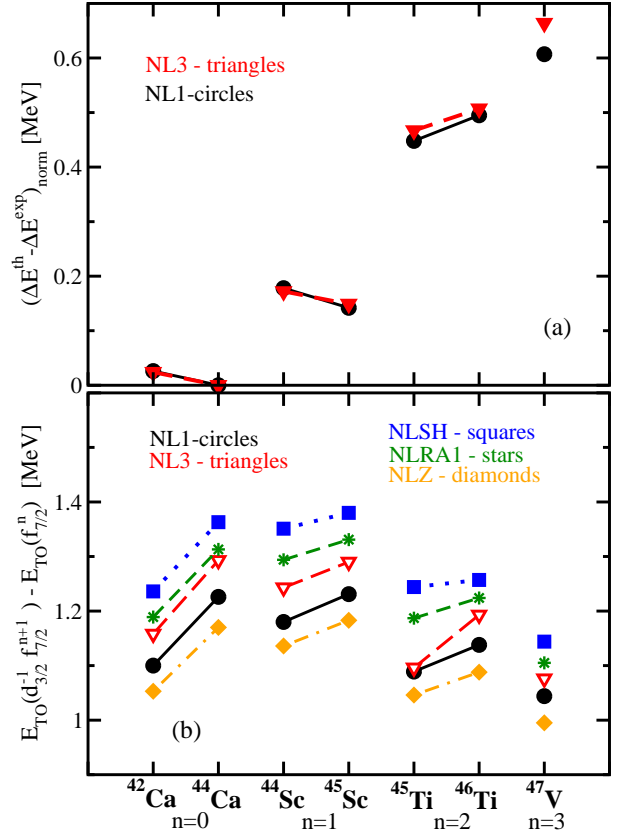}
\caption{\label{fig-exp-to} (Color online) (a) The difference 
$(\Delta E^{th}-\Delta E^{exp})_{norm}$ between the calculated and 
experimental values of $\Delta E$ (based on the results of Fig.\ 
\ref{excit}). This difference is normalized to zero  for $^{44}$Ca. 
(b) The calculated difference 
$E_{TO}(d_{3/2}^{-1}\,f_{7/2}^{n+1}) - E_{TO}(f_{7/2}^n)$ shown as a 
function of $n$ for the indicated RMF parametrizations (based on 
the results of Fig.\ \ref{fig-nm}).}
\end{figure}

  Fig.\ \ref{excit} compares the results of calculations with experiment. The 
same data set as in Refs.\ \cite{ZSW.05,BWSMG.06} is used in this comparison, but 
it is shown as a function of $n$,  where $n$ stands for the number of the $f_{7/2}$ 
protons in the $f_{7/2}^n$ terminating state. In addition, this figure compares 
the absolute values and not the differences between experimental results and 
calculations as it was done in Refs.\ \cite{ZSW.05,BWSMG.06}: the differences 
normalized to $^{44}$Ca are compared in Fig.\ \ref{fig-exp-to}a. Since the 
$n$ value is the same for each isotope chain ($n=0$ for the Ca isotopes, $n=1$ 
for the Cs isotopes, $n=2$ for the Ti isotopes, and $n=3$ for the V isotopes), 
the isospin dependence of $\Delta E$ is clearly visible. Different 
isotope chains show different isospin dependencies and they are well reproduced 
in the calculations (Fig.\ \ref{excit} and Fig.\ \ref{fig-exp-to}a). On the 
other hand, the calculations overestimate the absolute value of $\Delta E$ (Fig.\ 
\ref{excit}), and the difference between the calculated and experimental 
$\Delta E$ values show pronounced $n$-dependence (Fig.\ \ref{fig-exp-to}a).

  One source of the discrepancy between theory and experiment is related 
to the impact of effective mass of the nucleon on the single-particle spectra 
(see Sect.\ \ref{energy-scale}): in general, it should lead to an overestimate 
of experimental $\Delta E$ in the calculations.  The other sources of these 
discrepancies are analyzed in detail in this Section.
 
\subsection{Deformation polarization effects}

   The deformation polarization effects discussed in Sect.\ \ref{pol-eff} 
are characterized by the $\Delta E^{def-pol}$ energies. The sign of 
$\Delta E^{def-pol}$ depends on relative deformations of the $f_{7/2}^n$ 
and $d_{3/2}^{-1} f_{7/2}^{n+1}$ terminating states (Fig.\ \ref{def}). It 
is positive (negative) when the $\beta_2$-deformation of the 
$d_{3/2}^{-1} f_{7/2}^{n+1}$ state is larger (smaller) than the one of 
the $f_{7/2}^n$ state. The $\Delta E^{def-pol}$ values almost do not 
depend on the parametrization of the RMF Lagrangian: the difference in 
their values is below 10 keV if the results of the NL1 and NL3 parametrizations 
are compared. If the experimental data are corrected for these deformation 
polarization effects, then smooth trend (the curve 'exp. (def-cor)' in 
Fig.\ \ref{excit}) as a function of $n$ emerges. The $\Delta E$ value along 
this curve decreases by $\sim 0.25$ MeV on going from $n=0$ to $n=3$. 
Assuming that these effects are reasonably well described in the calculations, 
one can conclude that the nucleus-dependent fluctuations  in experimental 
value of $\Delta E$ (the curve 'exp.' in Fig.\ \ref{excit}) are due to 
deformation polarization effects.

\subsection{Nuclear magnetism (time-odd mean fields) in the terminating 
states of the $A\sim 44$ mass region}
\label{NM-A44}

\begin{figure}[h]
\includegraphics[angle=0,width=8cm]{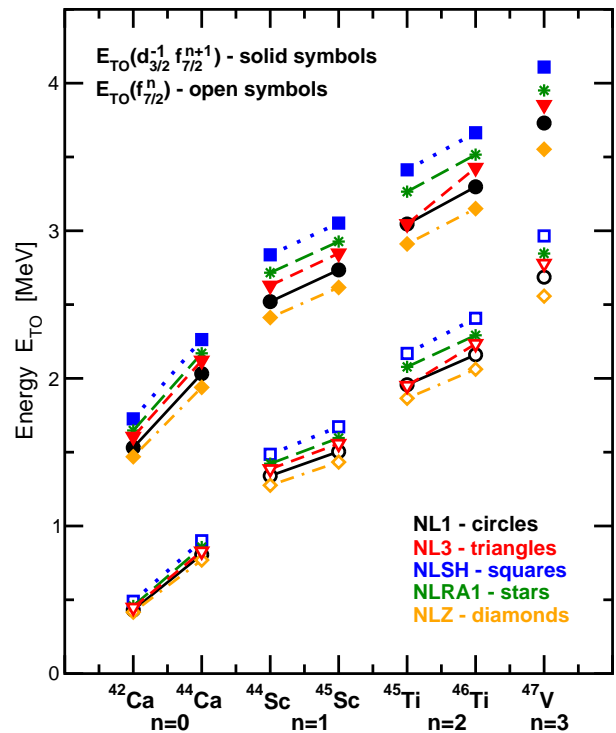}
\caption{\label{fig-nm} (Color online) Additional bindings $E_{TO}(state)$ 
(in absolute value) to the energies  of terminating states  due to NM 
shown for the terminating states of interest. The results are shown for the indicated 
parametrizations of the RMF Lagrangian as a function of $n$.}
\end{figure}

  Fig.\ \ref{fig-nm} shows the additional bindings $E_{TO}(state)$ to the energies 
of terminating states due to NM. This quantity 
increases with the increase of the $n$-value  for the $f_{7/2}^n$ and $d_{3/2}^{-1}\,f_{7/2}^{n+1}$ 
terminating states. The increase of $E_{TO}$ with isospin within specific isotope chain 
is associated with the increase of the number of the occupied neutron $f_{7/2}$ states
and corresponding increase in spin. The increase of the values of $E_{TO}$ correlates 
with the increase of the spin of the terminating states: for example, the $f_{7/2}^n$ 
and $d_{3/2}^{-1}\,f_{7/2}^{n+1}$ terminating states have $I_{max}=6^+$ and $I_{max}=11^-$ 
in $^{42}$Ca and  $I_{max}=\frac{31}{2}^-$  and $I_{max}=\frac{35}{2}^+$ in $^{47}$V, 
respectively.

  The results of the calculations confirm previous conclusion obtained in $^{20}$Ne (Sect.\ 
\ref{Ne20-sect}) that the additional binding due to NM is considerably enhanced in the 
terminating states. At no rotation, the additional binding due to NM to the energies 
of the single-particle configurations in odd-mass nuclei is in average around $\sim 100$ keV 
and seldom reaches 200 keV in the mass region of interest \cite{AA.08}. This is much smaller 
than the additional binding observed in the terminating states in which it reaches 4 MeV for 
$n=3$ (Fig.\  \ref{fig-nm}).

  Because of their magnitude, the $E_{TO}$ values in terminating states are also a 
good measure of how well the time-odd mean fields are defined in the specific 
version of DFT. The  $E_{TO}$ values obtained with different frequently used non-linear 
parametrizations of the RMF Lagrangian such as NL1 \cite{NL1}, NL3 \cite{NL3}, NLSH \cite{NLSH}, 
NLRA1 \cite{NLRA1} and  NLZ \cite{NLZ} are shown in Fig.\ \ref{fig-nm}.  With increasing 
$E_{TO}$ and $n$, the absolute variations in the $E_{TO}$ values calculated with different 
RMF parametrizations increase. However, they are still within 15\% of the absolute value of 
$E_{TO}$. This result suggests that within the non-linear versions of the RMF Lagrangian NM 
is defined with similar accuracy.

 This value can be used to estimate the uncertainty in the definition of the moments 
of inertia in the CRMF calculations due to the uncertainty in NM. Dependent on the nuclear 
system, the NM contribution to the total kinematic moment of inertia is approximately 10-25\%  
\cite{CRMF,AA.08}. Thus, the uncertainty of the definition of the absolute value of the 
total kinematic moments of inertia due to the uncertainty in the definition of NM is modest 
being in range of 1.5-3.75\%.

  It follows from Fig.\ \ref{fig-exp-to} that the portion of the $n$-independent part 
of the discrepancy between experimental and calculated $\Delta E$ values may be related 
to the uncertainties in NM since the difference $E_{TO}(d_{3/2}^{-1}\,f_{7/2}^{n+1}) - 
E_{TO}(f_{7/2}^n)$ somewhat (within $\approx 200$ keV) depends on the RMF parametrization.
The contribution of NM into the $n$-dependent part of this discrepancy
is discussed in Sect.\ \ref{n-depDE}.

\subsection{The dependence of $\Delta E$ on orbital angular momentum and
spin-orbit interaction}
\label{Dendep}

\subsubsection{The impact of density modifications on the single-particle properties}
\label{sp-impact}

\begin{figure}[h]
\includegraphics[angle=0,width=8cm]{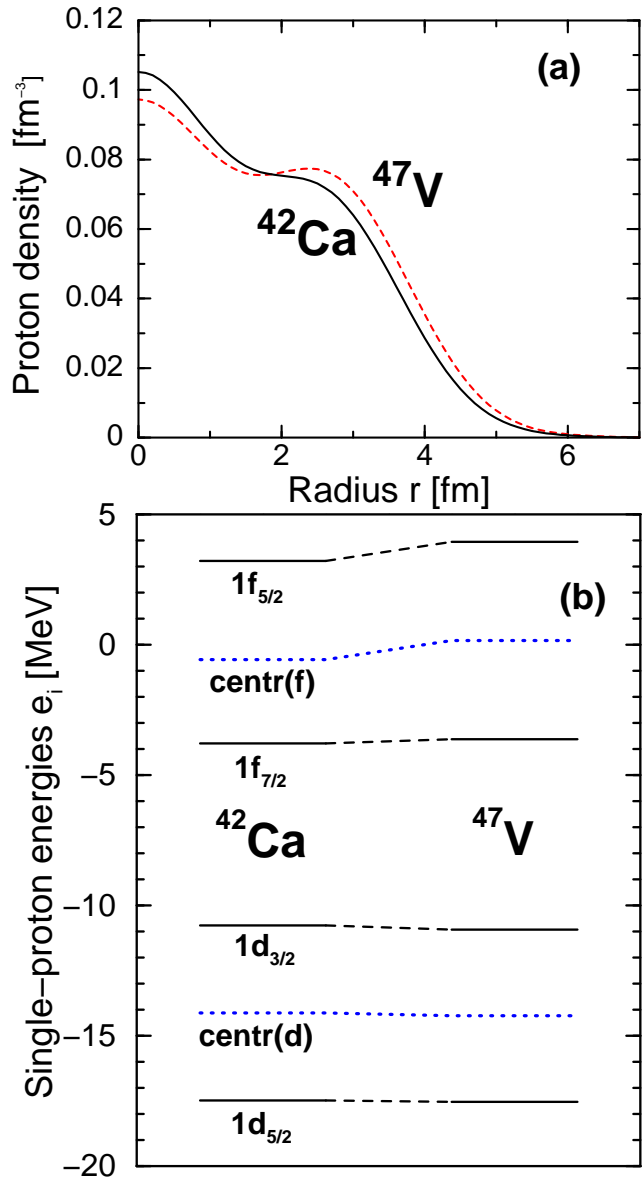}
\caption{\label{sp-ca-v} (Color online) (a) Proton density distribution in the $^{42}$Ca 
and $^{47}$V nuclei as obtained for ground state in the  spherical RMF 
calculations with the NL1 parametrization of the Lagrangian.  
(b) Corresponding energies of 
the single-particle states. The states in $^{42}$Ca are shown at the calculated 
energies, while all states in $^{47}$V are shifted by constant value in such 
a way that the average energy of the $1d_{3/2}$ and $1f_{7/2}$ states is the same as in $^{42}$Ca.}
\end{figure}

  It is well known fact that the modifications of the central nucleonic 
potential and its surface properties affect the single-particle states 
with different angular momentum {\it l} in a different way (see Refs.\ 
\cite{MBBD.85,RBRMG.98}  and references therein). They also alter the 
spin-orbit potential and lead to the changes in the spin-orbit splittings.  
In order to check how big this effect is in the nuclei under study, the proton 
density distributions and single-particle spectra at spherical shape are 
compared in Fig.\ \ref{sp-ca-v} for the $^{42}$Ca and $^{47}$V nuclei. These 
nuclei represent the lower and upper mass ends of the data set under investigation. 
The configuration of $^{47}$V has 2 additional $f_{7/2}$ neutrons and 3 additional 
$f_{7/2}$ protons as compared with the configuration of $^{42}$Ca. The 
filling of these high-$j$ orbitals increases the density near the surface (Fig.\ 
\ref{sp-ca-v}a). These modifications of the density change the central and 
spin-orbit nucleonic potentials (in a similar fashion as it was discussed in 
Refs.\ \cite{TPC.04,AF.05}) leading to the modifications of the single-particle 
spectra (Fig.\ \ref{sp-ca-v}b).

  On going from $^{42}$Ca to $^{47}$V, the spin-orbit splitting in the $d_{5/2}-d_{3/2}$ 
doublet decreases by 0.12 MeV (from 6.72 MeV to 6.60 MeV), while the one in the $f_{7/2}-f_{5/2}$ 
doublet increases by 0.57 MeV (from 7.0 MeV to 7.57 MeV). If the modifications in the 
single-particle spectra would be restricted only to the spin-orbit splittings and 
their modifications would evenly be redistributed between the $j=l+1/2$ and $j=l-1/2$ 
members of the spin-orbit doublet, this would decrease the $1f_{7/2} - 1d_{3/2}$ 
splitting by 0.22 MeV.

  However, the calculations show that the $f_{7/2}-d_{3/2}$ splitting is increased 
by 0.34 MeV (Figs.\ \ref{sp-ca-v}). Assuming that the energy scale does not change
on going from $^{42}$Ca to $^{47}$V, this can only be explained by the change of 
the relative positions of the $d$ (${\it l}=2$) and $f$ (${\it l}=3$) states 
from which the $d_{3/2}$ and $f_{7/2}$ states emerge. Unfortunately, there is no 
straightforward way in the RMF calculations to get an access to the $(N,l)$ states (in sense of 
Fig.\ \ref{method}). Thus, in order to illustrate the dependence on {\it l},  
the centroid energy (denoted as ``centr(state)'' in Fig.\ \ref{sp-ca-v}b) and 
defined  as an average energy of the members of spin-orbit doublet is used. 
  Fig.\ \ref{sp-ca-v}b shows that the energy gap between the centroids of the $d$ 
and $f$ spin-orbit doublets increases by 0.55 MeV on going from $^{42}$Ca to $^{47}$V. 
As a consequence of this increase and the above discussed changes in the spin-orbit 
splittings, the $1f_{7/2}-1d_{3/2}$ splitting increases by 0.34 MeV. This value 
represents more than half of the increase of $(\Delta E^{th}-\Delta E^{exp})_{norm}$ 
on going from $^{42}$Ca to $^{47}$V (Fig.\ \ref{fig-exp-to}a).

\subsubsection{Relative placements of the states with different angular momentum $l$}
\label{rel-plac}

  The fact that the relative placement of states with different orbital angular 
momentum $l$ (especially, of high-$l$ states) is not well reproduced in non-relativistic 
and relativistic mean field models is well known, see Refs.\ \cite{MBBD.85,RBRMG.98,LBBDM.07}. 
The origin of this problem is connected with the surface profile of the mean field and 
kinetic terms. Microscopic considerations indicate that the effective mass of 
the nucleon has a pronounced surface profile which is insufficiently parametrized in the 
present mean field models \cite{MBBD.85}. In Refs.\ \cite{ZSW.05,BWSMG.06}, this fact has 
been ignored and no proof has been provided that the placement of the $d$ and $f$ states, 
from which the $d_{3/2}$ and $f_{7/2}$ states, emerge is correct.

  It turns out that the difference in absolute value of $\Delta E$ obtained 
in the NL1 and NL3 parametrizations (Fig.\ \ref{fig-exp-to}) is well explained by the 
differences in the relative energies of the $d$  and $f$  
states in these parametrizations. The energy gap between the 
centroids of the $d$ and $f$ spin-orbit doublets is larger in the NL3 
parametrization as compared with the NL1 one by approximately 400 keV. If one 
corrects the NL1 results by this energy gap, one gets the results indicated as 
NL1$^{cor}$ in Fig.\ \ref{excit}. The NL1$^{cor}$ results are very close to the 
NL3 ones, which strongly suggests that the difference between the  NL1 and NL3 
results is predominantly due to different relative energies of the $l=3$ and 
$l=2$ states in these parametrizations of the RMF Lagrangian.

   Ref.\ \cite{BWSMG.06} has attributed the fact that the NL1 and NL3 
parametrizations differ in the description of the absolute $\Delta E$ value 
(Fig.\ \ref{excit}) to the magnitude  of the iso-scalar spin-orbit potential. 
The current investigation does not support this interpretation.

    These results suggest that instead of readjusting the isoscalar strength 
of the spin-orbit interaction as it was done in Ref.\ \cite{ZSW.05}, one 
can attempt to readjust the coupling constants of the DFT terms influencing 
the relative energies of the $l=2$ and $l=3$ states with the same effect 
on $\Delta E$. Indeed, 5\% reduction  of the isoscalar strength of the 
spin-orbit interaction introduced in Ref.\ \cite{ZSW.05} reduces $\Delta E$ 
by $\sim 350$ keV and this change in $\Delta E$ almost does not depend on 
nucleus (Fig. in Ref.\ \cite{SW.08}). On the other hand, the CRMF results
suggests that the same effect can be achieved if the relative distance
of the $d$ and $f$ states is modified. Indeed, the $\Delta E$ values 
obtained in the CRMF calculations decrease by approximately the same 
amount on going from the NL3 to NL1 parametrization of the RMF Lagrangian 
and this decrease only weakly depends on the nucleus (Fig.\ \ref{excit}).

\subsubsection{The $n$-dependence of $(\Delta E^{th}-\Delta E^{exp})_{norm}$}
\label{n-depDE}

\begin{figure}[h]
\includegraphics[angle=-90,width=8.5cm]{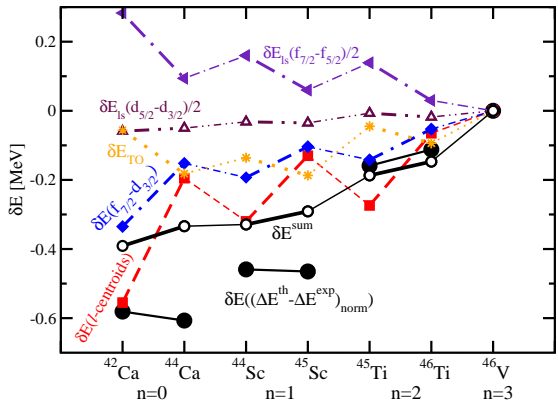}
\caption{\label{delta} (Color online) The variations $\delta E_i$ of different terms
contributing to the $\delta(\Delta E^{th}-\Delta E^{exp})_{norm})$ variation, see text 
for detail.}
\end{figure}

 The $(\Delta E^{th}-\Delta E^{exp})_{norm}$ quantity  shows pronounced 
dependence on $n$ (Fig.\ \ref{fig-exp-to}a) and its trend (if normalized 
to a single nucleus) almost does not depend on the RMF parametrization. 
In order to understand which ingredients of $\Delta E^{SC}$ contribute 
into this $n$-dependence, the variations $\delta E_i=\Delta E_i(nucleus)- \Delta E_i(^{47}V)$ 
of different ($i$-th) terms contributing to $\Delta E^{SC}$ are studied below. Contrary to Fig.\ 
\ref{fig-exp-to}a, $^{47}$V is selected as a reference in order to get a 
picture less disturbed by large fluctuations of some variations in the 
vicinity of $^{42}$Ca (Fig.\ \ref{delta}). The 
$\delta E_{TO}=\delta(E_{TO}(d_{3/2}^{-1}\,f_{7/2}^{n+1}) - E_{TO}(f_{7/2}^n))$
variation is obtained in the deformed CRMF calculations (from Fig. \ref{fig-exp-to}b), 
while other variations shown in Fig.\ \ref{delta} are calculated in spherical RMF 
calculations.  Thus, I effectively employ the approximation given in Eq.\ 
(\ref{eq-poleff}) assuming that the deformation polarization effects are the 
same both in theory and experiment. All the results presented here are based 
on the calculations with the NL1 parametrization, but it was checked that 
the NL3 results are similar.

   The $\delta((\Delta E^{th}-\Delta E^{exp})_{norm})$ variation indicates
that the difference between the calculated and experimental $\Delta E$ values
decreases with decreasing $n$. Note that for a given $n$ it is almost
constant indicating only weak isospin dependence of this variation.
The largest changes as a function of nucleus amongst the calculated 
variations are seen in the energy gap between the centroids of the $d$ 
and $f$ spin-orbit doublets (the curve denoted as ``$\delta E ({\it l}-{\rm centroids})$''
in Fig.\ \ref{delta}). It has the same trend as $\delta((\Delta E^{th}-\Delta E^{exp})_{norm})$ 
as a function of $n$. For a given $n$, it shows very large dependence on isospin. 
The second largest variation is seen in the spin-orbit splitting of the 
$f_{7/2}-f_{5/2}$ spin-orbit doublet (the curve denoted as ``$\delta E_{\it ls}(f_{7/2}-f_{5/2})/2$'' 
in Fig.\ \ref{delta}). The factor 1/2 is used in this variation since only one half of 
the total variation of spin-orbit splitting contributes into the $f_{7/2}-d_{3/2}$ splitting
(see Sect.\ \ref{sp-impact}). The $\delta E_{\it ls}(f_{7/2}-f_{5/2})/2$ variation has the 
wrong trend as compared with $\delta ((\Delta E^{th}-\Delta E^{exp})_{norm})$. The
$\delta E_{\it ls}(d_{5/2}-d_{3/2})/2$ variation of the spin-orbit splitting in the 
$d_{5/2}-d_{3/2}$ doublet is quite small. It has the correct trend as compared with 
$\delta((\Delta E^{th}-\Delta E^{exp})_{norm})$.

  The $\delta E(f_{7/2}-d_{3/2})$ variation of the $f_{7/2}-d_{3/2}$ splitting approximately 
satisfies the relation
\begin{eqnarray}
\delta E(f_{7/2}-d_{3/2})  =  \delta E({\it l}-{\rm centroids}) +  \nonumber \\ 
\delta E_{\it ls}(f_{7/2}-f_{5/2})/2  +  \delta E_{\it ls}(d_{5/2}-d_{3/2})/2
\end{eqnarray}
 The isospin dependencies seen in $\delta E ({\it l}-{\rm centroids})$ and 
$\delta E_{\it ls}(f_{7/2}-f_{5/2})/2$ act is opposite directions, thus, reducing the 
isospin dependence of $\delta E(f_{7/2}-d_{3/2})$ as compared with the one of 
$\delta E ({\it l}-{\rm centroids})$. However, the $\delta E(f_{7/2}-d_{3/2})$ variation  
(Sect.\ \ref{sp-impact}) cannot completely account neither for absolute value 
nor for isospin dependence (for a given $n$) of the $\delta((\Delta E^{th}-\Delta E^{exp})_{norm})$ 
variation.

   Only when the $\delta E(f_{7/2}-d_{3/2})$ variation is combined with 
the $\delta E_{TO}$ variation due to NM by
\begin{eqnarray}
\delta E^{sum} = \delta E(f_{7/2}-d_{3/2}) + \delta E_{TO}
\end{eqnarray}
a better description of the $\delta((\Delta E^{th}-\Delta E^{exp})_{norm})$ 
variation emerges.  For a given $n$, the isospin dependence of the 
$\delta((\Delta E^{th}-\Delta E^{exp})_{norm})$ is well described 
by $\delta E^{sum}$.  The absolute value of 
$\delta((\Delta E^{th}-\Delta E^{exp})_{norm})$ for the Ti nuclei 
is well described by $\delta E^{sum}$. However, for the Ca and 
Sc nuclei, the difference between these two quantities reaches
$30\%$ of the absolute value of $\delta((\Delta E^{th}-\Delta E^{exp})_{norm})$. 
The part of this discrepancy is definitely related to the limitations 
of the approximation given by Eq.\ (\ref{eq-poleff}).

  Thus, the current study clearly shows that the modifications of the 
relative placement of the states with different angular momentum $l$, 
the spin-orbit splittings and time-odd mean fields on
going from $^{47}$V to $^{42}$Ca contribute into the $n$-dependence 
of the difference between the calculated and experimental $\Delta E$
values (the $\Delta E^{th}-\Delta E^{exp})_{norm}$ quantity).
Previously, this $n$-dependence of $(\Delta E^{th}-\Delta E^{exp})_{norm}$,
expressed in a different form (Fig. 1 in Ref.\ \cite{BWSMG.06}, has been 
solely attributed to the deficiency of the iso-vector term of the
spin-orbit interaction \cite{BWSMG.06}, but the current investigation 
does not support such an interpretation.

\section{The energy scale and the effective mass of the nucleon}
\label{energy-scale}

  The terminating states are expected to be of predominantly single-particle 
nature \cite{CNS,ZSNSZ.07}: the $d_{3/2}^{-1}\,f_{7/2}^{n+1}$ terminating states are 
obtained from the $f_{7/2}^{n}$ terminating state by particle-hole (p-h) 
excitation from the $d_{3/2}$ state into the $f_{7/2}$ state. The energy of 
this p-h excitation depends on the energies of above mentioned states, 
and, thus, it is affected by the energy scale of the single-particle spectra 
which is related to the effective mass $m^*(k_F)/m$ of the nucleon at the 
Fermi surface.

 In the RMF theory, the spin-orbit interaction is effectively scaled by 
the effective mass of the nucleon (Ref.\ \cite{R.89}), and that is a reason 
why experimental data on spin-orbit splittings are well described in the calculations 
\cite{RBRMG.98,BRRMG.99}. 
This scaling also explains why the spin-orbit splittings of the $1p_{3/2}-1p_{1/2}$, 
$1d_{5/2}-1d_{3/2}$ and $1f_{7/2}-1f_{5/2}$ spin-orbit partner orbitals are 
almost the same in the RMF and the Nilsson potential calculations (Fig.\ 
\ref{single}). Note, that the Nilsson potential is characterized by the effective 
mass $m^*(k_F)/m \sim 1$ which is typical for experimental density of the
quasiparticle states.  Only in the case of the $2p_{3/2}-2p_{1/2}$ 
doublet, the spin-orbit splitting is smaller in the RMF calculations. 

  On the other hand, the energies of the centroids of the spin-orbit 
doublets are stretched out in the RMF calculations as compared with 
the Nilsson potential: the difference between the  RMF and Nilsson 
centroid energies increases on going away  from the Fermi level
(Fig.\ \ref{single}). Thus, the stretching out of the single-particle 
spectra due to low effective mass of the nucleon shows up mostly for 
orbital motion of particles and affects the relative placement of the 
levels  with different angular momentum {\it l}. The origin of this 
problem has been discussed in Sect.\ \ref{rel-plac}.

The effective mass of nucleon at the Fermi surface (Lorentz mass in the notation 
of Ref.\ \cite{JM.89} for the case of the RMF theory) is $m^*(k_F)/m \sim 0.65$ 
in the RMF theory \cite{BRRMG.99}, $\sim 0.7$ in the case of the Hartree-Fock 
(HF) approach based on the Gogny forces \cite{BHR.03}, and varies in the range 
$0.6-1.0$ in the HF approach based on the Skyrme forces \cite{BHR.03} revealing 
much larger flexibility of this type of the DFT with respect of effective 
mass. As a consequence of low effective mass, the calculated spectra are less 
dense than the experimental ones: the well known fact in non-relativistic and 
relativistic models both for spherical \cite{MBBD.85,LR.06,RBRMG.98} and deformed 
systems \cite{A250,BBDH.03}. This study shows that the $\Delta E^{SC}$ quantity 
differs from the $f_{7/2}-d_{3/2}$ energy gap in the spherical single-particle 
spectra only by the effects of time-odd mean fields and deformation polarization 
effects (Sect.\ \ref{pol-eff}). These facts 
pose an open problem on how to compare the experimental data on $\Delta E$ with the 
results of the DFT calculations (especially, those with low effective mass) since 
the experimental data on $\Delta E^{exp}$ is expected to be characterized by  
$m^*(k_F)/m \sim 1$. The implicit assumption used in Refs.\ \cite{ZSW.05,BWSMG.06} 
that the DFT reproduces the empirical $\Delta E$ values relatively well, say within 
$\sim 10\%$ \cite{SW.08}, may be too optimistic especially for the DFT with low 
effective mass. 

\section{Conclusions}
\label{concl}

   In conclusion, the following results were obtained in the study of band
termination within the DFT framework:

\begin{itemize}
\item
 At band termination, the NM does not modify neither total angular 
momentum nor the expectation values of the single-particle angular 
momenta  $<j_x>_i$ of the single-particle orbitals. NM provides an 
additional binding to the energies of the specific configuration 
and this additional binding increases with spin and has its 
maximum exactly at the terminating state. This suggests that the 
terminating states can be an interesting probe of the time-odd 
mean fields related to NM {\it provided that other effects can 
be reliably isolated.}

\item
  The realization of the TS-method in Refs.\ \cite{ZSW.05,BWSMG.06} is 
based on the analogy with simple form of the Nilsson potential which 
allows to neglect the deficiences in the relative placement of the states 
with different angular momentum $l$. This approximation is not valid for
terminating states in the $A\sim 44$ mass region in modern and 
most frequently used versions of the Nilsson potential 

\item
   The impact of the relative placement of the states with different
angular momentum $l$ on $\Delta E^{SC}$ is also clearly visible in 
the RMF calculations. The difference in absolute $\Delta E^{SC}$ values 
obtained in the CRMF calculations with the NL1 and NL3 parametrizations 
of the Lagrangian is defined by the different relative energies of the 
$l=3$ and $l=2$ states in these parametrizations. The modifications 
of the relative distance of the states with different angular momentum  
$l$  on going from $^{47}$V to $^{42}$Ca contribute 
into the $n$-dependence of the difference between the calculated and 
experimental $\Delta E$ values (the $\Delta E^{th}-\Delta E^{exp})_{norm}$ 
quantity) in addition to the ones due to the spin-orbit interaction
and time-odd  mean fields.

\end{itemize}

The detailed analysis of the TS-method in the RMF framework reveals 
the picture which is more complicated than the one suggested in Refs.\ 
\cite{ZSW.05,BWSMG.06}.  The relative placement of the states with 
different angular momentum ${\it l}$, defined by the properties of 
central potential, has to be taken into account in addition to the 
DFT terms discussed in these references when the $\Delta E$ quantity
is analyzed. Considering the similarities of the RMF theory and SDFT,  
it is very likely that these conclusions are also valid in the SDFT 
framework. The current investigation calls for a detailed study of  
the impact of the relative placement of the states with different 
orbital angular momentum ${\it l}$ on the $\Delta E^{SC}$ quantity
in the SDFT framework.

 Existing results for superdeformed bands in $^{32}$S \cite{MDD.00,Pingst-A30-60}
and low-spin states in odd mass nuclei \cite{AA.08} point to the time-odd mean 
fields as a major point of the difference between SDFT and RMF.  For example, 
the additional binding due to time-odd mean fields and the energy separation 
between different signatures of the SD bands are considerable stronger in
SDFT as compared with RMF \cite{MDD.00}. The current study clearly shows that the 
correlations induced by time-odd mean fields are large: additional binding due 
to NM reaches 4 MeV for $n=3$ (Fig.\ \ref{fig-nm}), which is by order of magnitude 
larger than those seen before in the RMF calculations at low spin. It also has 
a considerable impact on $\Delta E^{SC}$: $\Delta E_{TO} \sim 1.2$ MeV (Fig.\ 
\ref{fig-exp-to}b). These results call for a comparative study of time-odd mean
fields in the Skyrme DFT and RMF frameworks. Such study is necessary in order
to make a significant progress towards a better understanding of the role
of time-odd mean fields. The work in this direction is in progress and the
results will be presented in a forthcoming manuscript \cite{AA.08}.

\section{Acknowledgements}

The material is based upon work supported by the Department of Energy 
under Award Number DE-FG02-07ER41459.

\end{document}